\documentstyle[prl,aps]{revtex}
\begin{document}
\draft   

\title{Heavy-Exotic-Leptons at LHC}
\author{ J.\ E.\ Cieza Montalvo $^*$ and
P. P. de Queiroz Filho $^\dagger$}
\address{{\it Universidade do Estado do Rio de Janeiro,
Instituto de F\'{\i}sica, }\\
{\it 20559-900 Rio de Janeiro, RJ, Brazil}}


\maketitle

\begin{abstract}                  
We investigate the production, signatures and backgrounds of a pair of heavy-exotic charged leptons and a heavy-exotic neutrino together with the exotic charged lepton in the framework of the vector singlet model, vector doublet model, and fermion-mirror-fermion model, at the CERN Large Hadron Collider (LHC). We still show that the cross-sections for the case of the production of a pair of heavy exotic leptons are competitive with the ones for the single production of exotic leptons. 
\end{abstract}
\vskip 1pc]
\bigskip


\vskip 2.5cm

{\em Submitted to Phys.\ Lett.\ B}

\newpage



The standard electroweak theory (SM) provides a very satisfactory description of most elementary particle phenomena up to the presently available energies. However, the SM has many problems: the spectrum of elementary fermions, no unifications of the forces, CP violation and the existence of many parameters. Concerning the first problem, there is no theoretical explanation for the existence of several generations and for the values of the masses.  It was established at the CERN $e^{+} e^{-}$ collider LEP that the number of light neutrinos is three \cite{lep}. However, this result does not rule out the possibility that additional generations exist in which the masses of both the neutrino and the charged lepton exceed half of the mass of the $Z^{0}$ \cite{nov}.

Many models, such as composite models \cite{af,bu1}, grand unified theories \cite{la}, technicolor models \cite{di}, superstring-inspired models \cite{e6}, mirror fermions \cite{maa}, predict the existence of new particles with masses around the scale of $1$ TeV and they consider the possible existence of new generations of fermions.

In this work, we shall consider the possibility of existence of a new generation of leptons and analyze the production of a pair of exotic heavy 
leptons and a exotic heavy neutrino together with the exotic lepton,  through the Drell-Yan mechanism at  the large hadron collider LHC ($\sqrt{s} = 14$ TeV). We also present here another form to produce these leptons, namely through the reactions $pp \rightarrow \gamma \gamma \rightarrow  L^{-} L^{+}$ and $pp \rightarrow Z \gamma \rightarrow  L^{-} L^{+}$, moreover we show that the production of a pair of exotic leptons is dominant over the single heavy lepton production \cite{sim98}, that is so, due to the vector component of the Z-boson coupling to exotic leptons and to the exchange of a photon in the $s$ channel (which does not appear in the case for the single heavy lepton production), although the phase space is more reduced for pair production than for single heavy lepton production.

The three models that we consider here include new fermionic degrees of freedom, which naturally introduce a number of unknown mixing angles and fermionic masses \cite{tom}. These models are: the vector singlet model (VSM) \cite{gon}, vector doublet model (VDM) \cite{riz} and the fermion-mirror-fermion model (FMFM) \cite{maa}.

Exotic leptons mixed with the standard leptons interact through the
standard weak vector bosons $W^{+}, W^{-}$ and $Z^{0}$, according to the interacting Lagrangians

\begin{equation}
{\cal L}_{\rm NC} = \frac{g}{4 cos\theta_{W}} \left[{ \bar{F}_{i}}
\gamma^{\mu} (g_{V}^{ij} - g_{A}^{ij} \gamma^{5}) F_{j} +
{\bar{F}_{i}} \gamma^{\mu} (g_{V}^{ij} - g_{A}^{ij} \gamma^{5}) f_{j} \right]
Z_{\mu} 
\label{lag1}
\end{equation}
   
and 

\begin{equation}
{\cal L}_{\rm CC} = \frac{g}{2 \sqrt{2}} { \bar{L}^{0} \gamma^{\mu} 
(C_{V}^{NL} - C_{A}^{NL} \gamma^{5}) E W_{\mu} } ,
\label{lag2}
\end{equation}
where $g_{V}^{ij}$ and $g_{A}^{ij}$ are the corresponding neutral vector-axial coupling constants, which are given in \cite{cie,sim98}, and $C_{V}^{NL}$ and $C_{A}^{NL}$ are
the charged vector-axial coupling constants, given in Table I for each of the three models that we study here.

TABLE I. Coupling constants for a charged heavy fermion 
interaction: for the VSM, the VDM and the FMFM: 

\vskip 1cm

\begin{tabular}{|c|c|c|c|}  \hline\hline
Cou. & VSM & VDM & FMFM \\ \hline  $C_{V}^{NL}$ & $\sin \theta_{L}^{\nu}  \sin\theta_{L}^{e}$ & $\cos (\theta_{L}^{\nu} - \theta_{L}^{e}) - 
\cos \theta_{R}^{\nu} \cos \theta_{R}^{e}$ & $\sin \theta_{L}^{\nu} 
\sin \theta_{L}^{e} + \cos \theta_{R}^{\nu} \cos \theta_{R}^{e}$ \\
\hline $C_{A}^{NL}$ & $\sin \theta_{L}^{\nu} \sin \theta_{L}^{e}$ 
& $\cos (\theta_{L}^{\nu} - \theta_{L}^{e}) + \cos \theta_{R}^{\nu} 
\cos \theta_{R}^{e}$ & $\sin \theta_{L}^{\nu} \sin \theta_{L}^{e} - 
\cos \theta_{R}^{\nu} \cos \theta_{R}^{e}$ \\ \hline 
\end{tabular} 

\vskip 0.6cm

We consider here that all mixing angles have the same value $\theta_{i}
= 0.1$, although phenomenological analysis \cite{tom1} gives an upper
bound of $sin^{2} \theta_{i} \leq 0.03$. This means that the value of
$\theta_{i}$ can be scaled up to $0.173$.

We begin our study with the mechanism of the Drell-Yan production of a pair of exotic charged leptons, that is we study the process $pp \rightarrow q   \bar{q} \rightarrow L^{-} L^{+}$. It takes place through the exchange of a $Z^{0}$ and $\gamma$ bosons in the $s$ channel. 

Using the interaction Lagrangians, Eqs. ($1$) and ($2$), we evaluate the cross section for the process $q \bar{q} \rightarrow L^{-} L^{+}$,  obtaining:
  
\begin{eqnarray} 
\left (\frac{d \hat{\sigma}}{d\cos \theta} \right )_{L^{+} L^{-}} = &&\frac{\beta \alpha^{2} \pi}{s^{2}} \Biggl [\frac{e_{q}^{2}}{s} ( 2 s M_{L}^{2} + (M_{L}^{2} - t)^{2} + (M_{L}^{2} - u)^{2}  )  \nonumber \\
&&+ \frac{e_{q}}{2 \sin^{2} \theta_{W} \cos^{2} \theta_{W} (s - M_{Z}^{2})} ( 2 s M_{L}^{2} g_{V}^{LL} g_{V}^{q}  \nonumber \\
&&+g_{V}^{LL} g_{V}^{q} ( (M_{L}^{2} - t)^{2} + (M_{L}^{2} - u)^{2}  )  + g_{A}^{LL} g_{A}^{q} ( (M_{L}^{2} - u)^{2} - (M_{L}^{2} - t)^{2} )  \Biggr ] \nonumber \\
&&+ \frac{\beta \pi \alpha^{2}}{16 s \cos^{4} \theta_{W} \sin^{4} \theta_{W}} \frac{1}{(s - M_{Z}^{2} )^{2}} \Biggl [(g_{V}^{{LL}^{2}} + g_{A}^{{LL}^{2}}) (g_{V}^{{q}^{2}} + g_{A}^{{q}^{2}}) ((M_{L}^{2} - u)^{2}  \nonumber \\
&&+(M_{L}^{2} - t)^{2}) + 2 (g_{V}^{{LL}^{2}} - g_{A}^{{LL}^{2}}) (g_{V}^{{q}^{2}} + g_{A}^{{q}^{2}}) s M_{L}^{2}   \nonumber \\
&&+ 4 g_{V}^{LL} g_{A}^{LL} g_{V}^{q} g_{A}^{q} ((M_{L}^{2} - u)^{2} - (M_{L}^{2} - t)^{2}) \bigr )   \Biggr] \; ,
\end{eqnarray}
where $\beta_{L} = \sqrt{1- 4 M_{L}^{2}/s}$ is the velocity of the 
exotic-fermion in the c.m. of the process, $M_{Z}$ is the mass of the $Z$ boson, $\sqrt{s}$ is
the center of mass energy of the $q \bar{q}$ system, $t = M_{L}^{2}
- \frac{s}{2} (1 - \beta \cos \theta)$ and {} $u = M_{L}^{2} - \frac{s}{2}
(1 + \beta \cos \theta)$, where $\theta$ is the angle between the exotic
lepton and the incident quark, in the c.m. frame.

The total cross section for the process $pp \rightarrow qq \rightarrow 
L^{-} L^{+}$ is related to the subprocess $qq \rightarrow L^{-} L^{+}$ 
total cross section $\hat{\sigma}$, through

\begin{equation}  
\sigma (s) =
\int_{\tau_{min}}^{1}
\int_{\ln \sqrt{\ \tau_{min}}}^{-\ln \sqrt{\ \tau_{min}}} d\tau \ dy \
q(\sqrt{\tau}e^y, Q^2) \ q(\sqrt{\tau}e^{-y}, Q^2) \ \hat{\sigma}(\tau, \hat{s})  \; , 
\end{equation}
where $\tau = \frac{\hat{s}}{s} (\tau_{min} = \frac{4 M_L^2}{s})$,
with $s$ being the center-of mass  energy  of the $pp$ system and $q(x,Q^2)$ is the quark structure function \cite{gluck}.

Another way to produce a pair of heavy exotic leptons is through the  elastic reaction of the type $pp \rightarrow \gamma \gamma  \rightarrow L^{-} L^{+}$, the inelastic $pp \rightarrow q \bar{q} \rightarrow \gamma \gamma  \rightarrow L^{-} L^{+}$ and the semielastic $pp \rightarrow Z \gamma  \rightarrow L^{-} L^{+}$ ones. The three processes take place through the exchange of the exotic lepton in the $t$ and $u$ channels. We will have, for the subprocess for the case of the elastic reaction, the following expression

\begin{eqnarray} 
\left (\frac{d \hat{\sigma}}{d\cos \theta} \right ) = &&\frac{\beta \alpha^{2} \pi}{s} \bigl [\frac{1}{(t - M_{L}^{2})^{2}} (-M_{L}^{4}- 3 M_{L}^{2} t - M_{L}^{2} u + tu)  \nonumber \\
&&+ \frac{1}{(u - M_{L}^{2})^{2}} (-M_{L}^{4}- M_{L}^{2} t - 3 M_{L}^{2} u + tu) \nonumber  \\ 
&&+ \frac{2}{(t - M_{L}^{2}) (u - M_{L}^{2})} (-2 M_{L}^{4} - M_{L}^{2} t - M_{L}^{2} u ) \bigr ]  \; ,
\end{eqnarray}
where $M_{L}$ is the mass of the exotic lepton and the other parameters are the same as above. For the subprocess for the case of the semielastic  reaction we also have the same expression as \cite{bhat}. The pair production cross section is then obtained by folding $\hat{\sigma}_{el,inel,semi}$ with the $\gamma$ ($f_{\gamma/p}, f_{\gamma/q}$) and Z ($f_{Z/q}$) distributions in the proton and quark, so that for the elastic production we have 

\[
\sigma_{el} (s) = \int_{\tau_{min}}^{1} d \tau \int_{\tau/x_{max}}^{x_{max}} \frac{dx_{1}}{x_{1}} \ f_{\gamma/p} (x_{1}) \ f_{\gamma/p} (\tau/x_{1}) \int_{-1}^{+1} \frac{d \hat{\sigma}}{dcos \theta} d cos \theta (\hat{s}).
\]
for the inelastic case we obtain

\begin{eqnarray} 
\sigma_{inel} (s) = &&\int_{\tau_{min}}^{1} d \tau \int_{\tau/x_{max}}^{x_{max}} \frac{dx_{1}}{x_{1}} \int_{x_{1}}^{1} \frac{d x_{3}}{x_{3}} \ f_{\gamma/q} (x_{3}) \ f_{q_{i}/p} \left (\frac{x_{1}}{x_{3}} \right )  \nonumber  \\
&& \int_{\tau/x_{1}}^{1} \ f_{\gamma/q} (x_{5}) \frac{d x_{5}}{x_{5}} f_{q_{j}/p} \left (\frac{\tau}{x_{1} x_{5}} \right ) \int_{-1}^{+1} \frac{d \hat{\sigma}}{dcos \theta} d cos \theta (\hat{s}),  \nonumber  
\end{eqnarray}  
and for the semielastic we have 

\[
\sigma_{semi} (s) = \int_{\tau_{min}}^{1} d \tau \int_{\tau/x_{max}}^{x_{max}} \frac{dx_{1}}{x_{1}} \ f_{\gamma/p} \left (\frac{\tau}{x_{1}} \right )  \int_{x_{1}}^{1} \frac{d x_{3}}{x_{3}} \ f_{Z/q} (x_{3}) \ f_{q/p} \left (\frac{x_{1}}{x_{3}} \right )   \int_{-1}^{+1} \frac{d \hat{\sigma}}{dcos \theta} d cos \theta (\hat{s})  \]

Now, the production of an exotic neutrino can be studied through the analysis of the charged current reaction of the type $pp \rightarrow q \bar{q}'  \rightarrow L^{+} N$, involving the  exchange of a $W^{\pm}$ boson in the $s$ channel. The subprocess cross section is 

\begin{eqnarray} 
\left (\frac{d \hat{\sigma}}{d\cos \theta} \right)_{LN} = &&\frac{\beta \alpha^{2} \pi |V_{pm}|^{2}}{16 \sin \theta^{4}_{W} s (s - M_{W}^{2})^{2}} \bigl [(M_{L}^{2} M_{N}^{2}- (s- t+ u)t)(g_{V}^{LN}+ g_{A}^{LN})  \nonumber \\
&&+ M_{L} M_{N} s ({g_{V}^{LN}}^{2} - {g_{A}^{LN}}^{2}) + (t- u) \frac{s}{2} (g_{V}^{LN}- g_{A}^{LN})^{2}
\bigr ]  \; ,
\end{eqnarray}
where $\beta$, t and u are, respectively, 

\[
\beta_{N,L} = \frac{\left [ \left (1- \frac{(M_{N} + M_{L})^{2}}{s} \right ) \left (1- \frac{(M_{N}- M_{L})^{2}}{s} \right) \right ]^{1/2}}{1 \pm \frac{M_{N}^{2}- M_{L}^{2}}{s}}, 
\]

\[
t= M_{N}^{2}- \frac{s}{2} \left [ \left(1+ \frac{M_{N}^{2}- M_{L}^{2}}{s} \right )  
- \left ( \left (1- \frac{(M_{N}+ M_{L})^{2}}{s} \right ) \left (1- \frac{(M_{N}- M_{L})^{2}}{s} \right ) \right )^{1/2} \cos \theta \right ],
\]

\[
u= M_{N}^{2}- \frac{s}{2} \left [ \left(1+ \frac{M_{N}^{2}- M_{L}^{2}}{s} \right )  
+ \left ( \left (1- \frac{(M_{N}+ M_{L})^{2}}{s} \right ) \left (1- \frac{(M_{N}- M_{L})^{2}}{s} \right ) \right )^{1/2} \cos \theta \right ],
\]
$V_{pm}$ are the matrix elements of CKM and $M_{W}$ is the mass of the gauge boson. Folding the proton structure functions with the elementary cross section above, we obtain

\begin{equation}  
\sigma_{NL}(s) =
\int_{\tau_{min}}^{1}
\int_{\ln{\sqrt{\tau_{min}}}}^{-\ln{\sqrt{\tau_{min}}}} d\tau \ dy \
q(\sqrt{\tau}e^y, Q^2) \ q(\sqrt{\tau}e^{-y}, Q^2) \ \hat{\sigma}_{NL}(\tau, s)  \; .
\end{equation}

The process for the heavy lepton production in hadronic colliders was well studied in the literature \cite{dic} and was shown that the dominant contributions are the well known Drell-Yan process and gluon-gluon fusion \cite{scot,dic}.

We present in Fig. $1$ the cross sections for the process 
$pp \rightarrow qq \rightarrow L^{-} L^{+}$, involving the three models considered here: VSM,  VDM and FMFM, for the LHC ($14$ TeV), using the mechanism of Drell-Yan. In all calculations we take $\sin^{2}_{\theta_W} = 0.2315$, $M_Z = 91.188$ and $M_{W} = 80.419$. For a planned LHC integrated luminosity of $ 10^{5} pb^{-1}/yr$ and taking the mass of the  lepton  equal to $200$ GeV, we have a total of $175 \cdot 10^2$ exotic lepton pairs produced per year for the VSM, $548 \cdot 10^2$ for the VDM and $684 \cdot 10^2$ for the FMFM.

In Figs. $2$ and $3$ we consider $L \bar{L}$ pair production with both $L$ and $\bar{L}$ decaying into leptons, that is, we have the   following subprocess 

\[
q \bar{q} \rightarrow Z, \gamma \rightarrow L \bar{L} \rightarrow  \bar{e}  e \bar{e} \ (\bar{\nu}_{e} \bar{e} \nu_{e} )  
\]

We initiate our analysis by considering that the signal for $L^{-} L^{+}$ pair production will be an electron, three positrons, neutrino and antineutrino, that is $e^{+} e^{-} e^{+} \ \bar{\nu}_{e} e^{+} \nu_{e}$ with the corresponding $Z^{0}$ and $W^{\pm}$ propagators. We compare this signal with the standard model background, such as $pp  \rightarrow ZZ$; this background can easily be distinguished and  therefore eliminated, by measuring the transverse mass of the $e^{-} e^{+}$ pair.

The numerical calculation for Figs. $2$ and $3$ was made in two stages, first, the total cross section for the process $p p \rightarrow q \bar{q}  \rightarrow Z, \gamma \rightarrow L \bar{L}$ was calculated, then the decay rate of the heavy leptons \cite{sim4} was used, and afterwards the product of both was taken. In these figures the distributions of  $p_{T}\!\!\!\!\!\!\slash$ \ and the transverse mass of the $e^{+} e^{-} e^{+}$ are showed.

In Fig. $4$ we show the cross section production of $L^{-} L^{+}$ for the case of the Drell-Yan, elastic, inelastic and semielastic cases, one can  see that the Drell-Yan case is three orders of magnitude greater than the others mechanisms, for $m_L=100$ GeV, and, for $m_L$ around $2$ TeV, the  $Z\gamma$ inelastic process is competitive with the Drell-Yan mechanism.  We can also observe that the elastic, inelastic and semielastic production in the FMFM is one order of magnitude greater than the results of Ref.  \cite{bhat}.

In Fig. $5$ it is shown the cross section production of exotic neutrinos together with the exotic charged lepton in comparison with the production of the pair of exotic charged leptons. It can be seen that both figures differ little one from another. The signal for $N L^{\pm}$ production will be five charged leptons and one neutrino, that is $e^{-} e^{+} \nu e^{-} e^{+} e^{-}$ with the $W^{\pm}$ and $Z^{0}$ propagators. By comparing this signal with the standard model background such as $pp \rightarrow WZZ$, we have that the $W Z Z$ production process has a clean experimental signature with $W^{+} \rightarrow e^{+} \nu$, $Z \rightarrow e^{-} e^{+}$ and $Z \rightarrow e^{-} e^{+}$ decays.

Figs. $6$ and $7$ show the distributions of  $p_{T}\!\!\!\!\!\!\slash$ \ \ and of transverse mass $M_{T} (e^{-} e^{+} \nu)$. We see in Fig. $7$ a Jacobian peak near the $M_{T} = M_{N}$. The numerical calculations of $M_{T}$ are made, in the same manner as above, which is defined in Ref. \cite{bar} by 

\begin{equation}
M_{T}^{2} (e^{-} e^{+} \nu) = \left [(p_{ee \nu T}^{2} + M_{ee \nu}^{2} )^{1/2} + p_{\nu T} \right ]^{2} - (\vec{p}_{eeT} + \vec{p}_{\nu T} )^{2} \; .
\end{equation}

The discovery of a pair of heavy-leptons will be without any doubt of great importance for the life beyond the standard model. In this work is analysed the signals and backgrounds for such particles and we can see, if we required the production of $100$ of such pairs per year for the LHC, we will be able to discover pairs of heavy-leptons with masses up to $900$ GeV. Since this number of events ($100$) is rather arbitrary, then it is interesting to know how the discovery limits of mass changes, when the number of events required for establishing the signal is varied. For example if we require 10(2000) events, the discovery limit of mass changes to 1360(350) GeV.   

\acknowledgments

We would like to thank to Prof. O. J. P. \'Eboli for calling the attention to some points.


\newpage

\begin{center}
FIGURE CAPTIONS
\end{center}

\vspace{0.5cm}

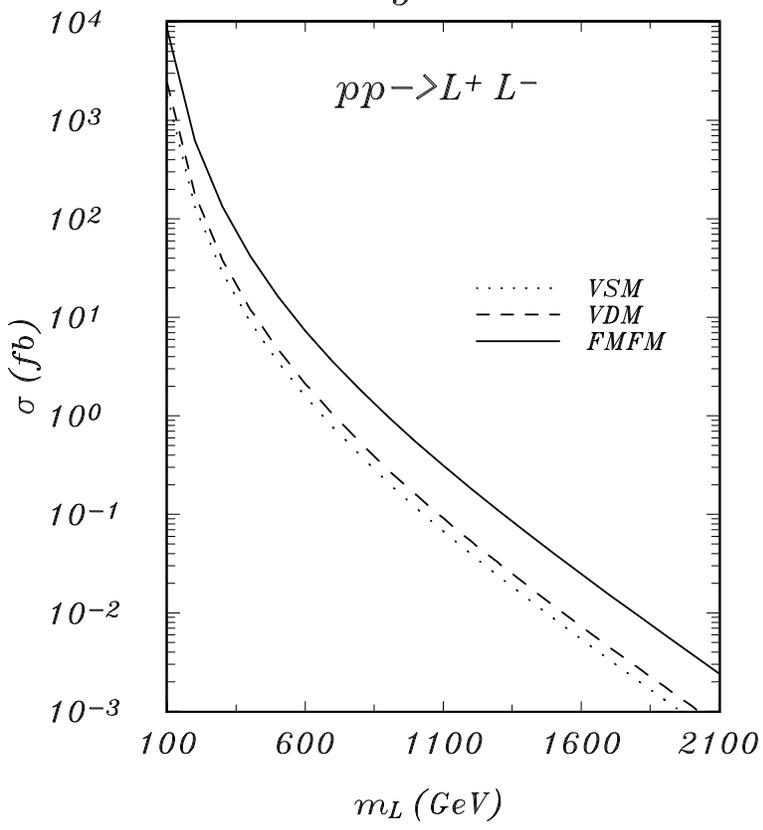
{\bf Figure 1}: Total cross section for the process $pp 
\rightarrow L^{+} L^{-}$ as a function of $M_{L}$ at $\sqrt{s} = 14$ TeV: 
(a) vector singlet model (dotted line); (b) vector doublet model
(dashed line); (c) fermion mirror fermion model (solid line).

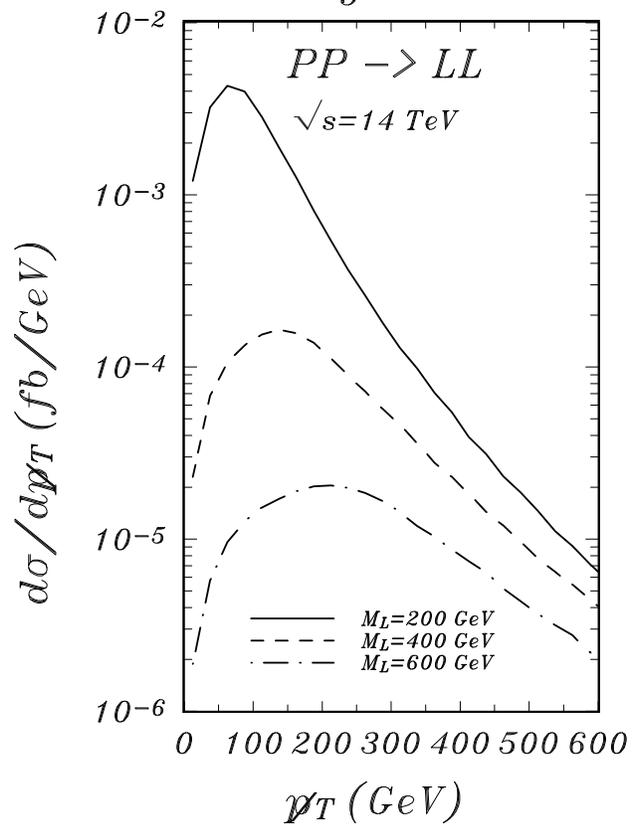
{\bf Figure 2}: Missing transverse momentum distributions from $L^{-} L^{+}$ production at $\sqrt{s} = 14$ TeV, for different exotic leptons mass: (a) $M_{L} = 600$ GeV  (dot-dashed line); (b) $M_{L} = 400$ GeV (dashed line); (c) $M_{L} = 200$ GeV (solid line).

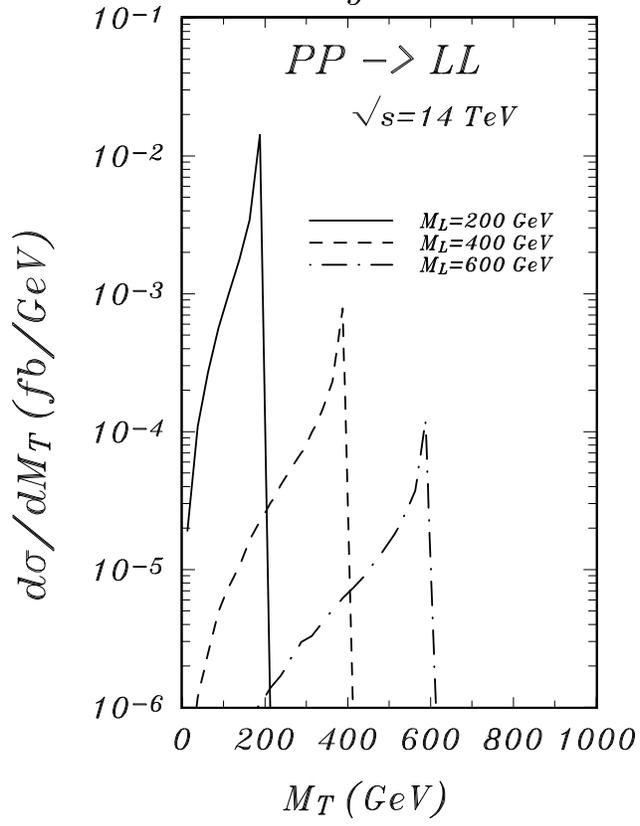
{\bf Figure 3}: Distribution of transverse mass $M_{T} (e^{+}, e^{-}, e^{+})$ from  $L^{-} L^{+}$ production at $\sqrt{s} = 14$ Tev: (a) $M_{L} = 600$ GeV (dot-dashed line); (b) $M_{L} = 400$ GeV (dashed line); (c) $M_{L} = 200$ GeV (solid line).

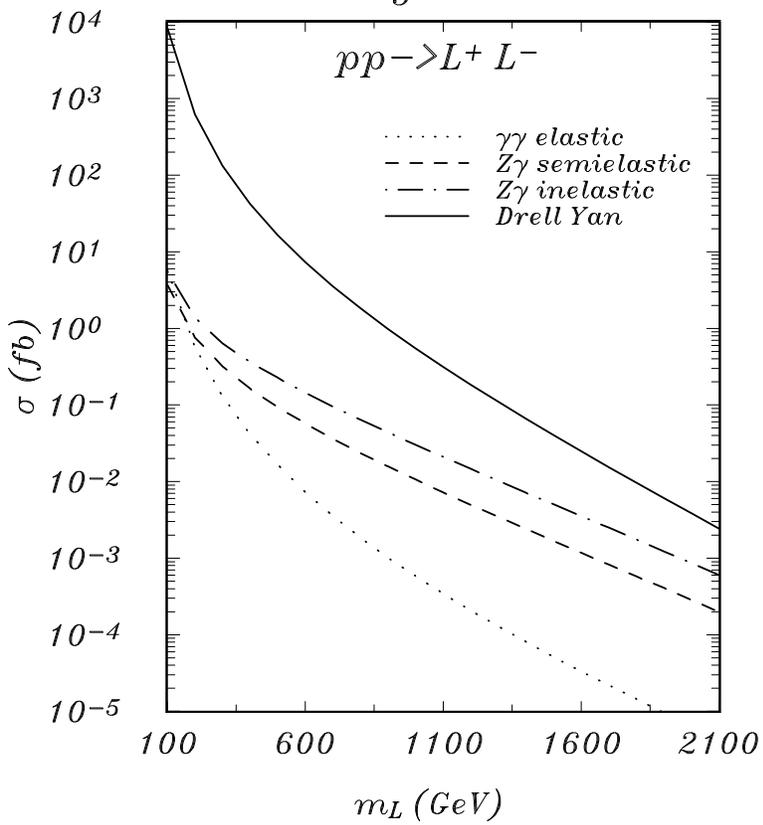
{\bf Figure 4}: Total cross section for the process $pp \rightarrow 
L^{+} L^{-}$ as a function of $M_{L}$ at $\sqrt{s} = 14$ TeV for different production mechanisms: (a) $\gamma\gamma$ elastic production (dotted line); (b) $Z\gamma$ semielastic production (dashed line); (c) $Z\gamma$ inelastic production (dot-dashed line); (d) Drell Yan production (solid line).

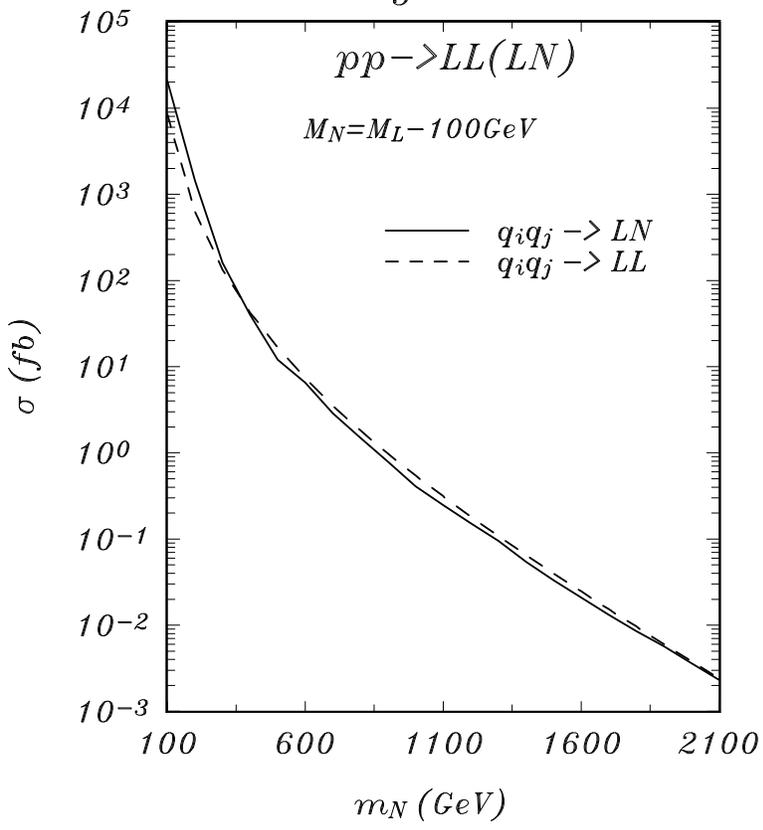
{\bf Figure 5}: Total cross section for the process $pp \rightarrow N L^{\pm}$ and $pp \rightarrow L^{-} L^{+}$ as a function of $M_{L}$ at $\sqrt{s} = 14$ TeV: (a) $N L^{\pm}$ production (dashed line); (b) $L^{-} L^{+}$ production (solid line).

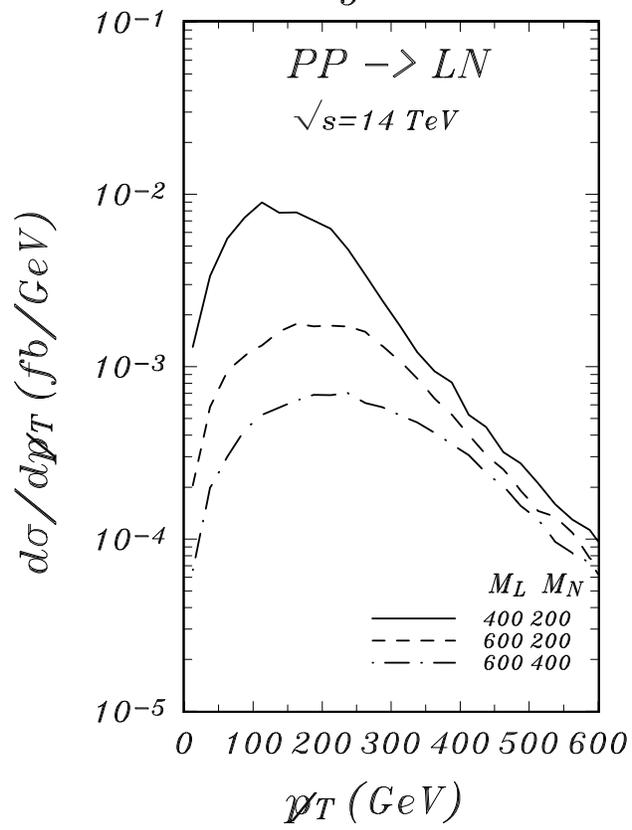
{\bf Figure 6}: Missing transverse momentum distributions from $N L^{\pm}$ production at $\sqrt{s} = 14$ TeV, for different exotic leptons mass: (a) $M_{L} = 600$ and $M_{N} = 400$ GeV  (dot-dashed line); (b) $M_{L} = 600$ and $M_{N} = 200$ GeV (dashed line); (c) $M_{L} = 400$ and $M_{N} = 200$  GeV (solid line).

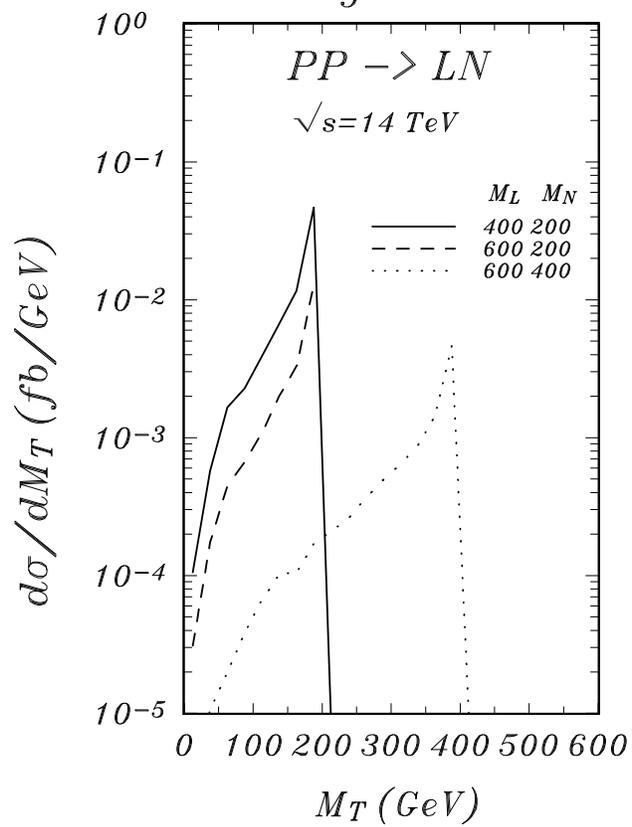
{\bf Figure 7}: Distribution of transverse mass $M_{T} (e^{\pm}, e^{\pm}, \nu(\bar{\nu}))$ from  $N L^{\pm}$ production at $\sqrt{s} = 14$ Tev: (a) $M_{L} = 600$ and $M_{N} = 400$ GeV (dotted line); (b) $M_{L} = 600$ and $M_{N} = 200$ GeV (dashed line); (c) $M_{L} = 400$ and $M_{N} = 200$ GeV (solid line).

\end{document}